\documentclass[useAMS,usenatbib]{mn2e}
\usepackage{graphicx}

\title[]{The Red Giant Branch in Near\--Infrared Colour\--Magnitude
Diagrams. II: The luminosity of the Bump and the Tip
\thanks{Based on data taken at the ESO-MPI 2.2m Telescope
	equipped with the near-IR camera IRAC2-ESO, La Silla (Chile).}}

\author[Valenti, Ferraro \& Origlia]{E. Valenti$^{1,2}$, F.~R. Ferraro$^1$, 
        L. Origlia$^{2}$    \\
 $^1$ Dipartimento Astronomia, Universit\`a di Bologna,  
      Via Ranzani 1, I-40127 Bologna, Italy ,\\
      e-mail elena.valenti2@studio.unibo.it, ferraro@bo.astro.it\\
 $^2$ INAF-Osservatorio Astronomico di Bologna, Via Ranzani 1, I-40127 Bologna,
      Italy, \\
      e-mail origlia@bo.astro.it \\
       }
\date{\today}

\begin{document}
\pagerange{\pageref{firstpage}--\pageref{lastpage}} \pubyear{2004}
\maketitle
\label{firstpage}

\begin{abstract}
We present new empirical calibrations of the Red Giant Branch (RGB)
Bump and Tip based on a homogeneous near\--Infrared database of 24 
Galactic Globular Clusters. 
The luminosities of the RGB Bump and Tip in the J, H and K bands 
and their dependence on the cluster metallicity have been 
studied, yielding empirical relationships.
By using recent transformations between the observational 
and theoretical planes, we also derived similar calibrations in terms
of bolometric luminosity. Direct comparison between updated theoretical
models and observations show an excellent agreement.
The empirical calibration of the RGB Tip luminosity in the near\--Infrared
passbands presented here is a fundamental tool to derive distances to far galaxies 
beyond the Local Group, in view of using the new ground-based adaptive optics facilities 
and, in the next future, the James Webb Space Telescope.
\end{abstract}


\section{Introduction}
This is the second in a series of papers aimed at studying
the Red Giant Branch (RGB) photometric properties and its
major evolutionary features in the near\--Infrared (IR) spectral 
domain.
In the first paper \citep[][ hereafter Paper~I]{pap1}
an extensive database, collected by our group over the last 10 years,
has been presented. By combining a new sample of 10 Galactic
Globular Clusters (GGCs) belonging to different galactic populations 
(i.e. Halo and Bulge) with the data set published by 
\citet[][ hereafter F00]{F00}, \citet{V04} and \citet[][ hereafter S04]{BB}, a 
homogeneous sample of 24 GGCs spanning a wide metallicity range
($-2.12{\leq}$[Fe/H]${\leq}$-0.49) has been obtained (see Table~\ref{par}).
In Paper~I, the entire database, calibrated in the 
{\it Two Micron All Sky Survey}
(2MASS) photometric system,
has been used to measure a set
of photometric indices describing the RGB: {\it i)~} location (colours and
magnitudes), and {\it ii)~} morphology (slope). 
Updated calibrations of these indices in terms of the clusters metallicity 
have been also derived. 
More details on the dataset can be found in Paper~I.\\
In the present paper the major RGB evolutionary features, namely the
Bump and the Tip, are measured in the J, H and K bands.
Their calibrations as a function of the clusters metallicity, in both the
observational and theoretical planes, are derived and discussed.\\
The RGB Tip luminosity is a bright standard candle and turns to be particularly 
useful to measure galaxy distances. The extension of the RGB Tip calibrations 
to the near\--IR passbands is of fundamental importance since the recent advent 
of ground-based adaptive optics systems and the future availability of  
the James Webb Space Telescope are allowing to spatially resolve bright giants 
and to accurately measure their magnitudes up to a few Mpc distances.\\
In \S 2 and \$ we measure the near\--IR magnitudes of the RGB Bump and Tip, 
respectively, of our sample of 24 GGCs and we investigate their behaviour with 
varying the cluster metallicity. 
The results of the transformations between the 
observational and theoretical planes for the RGB Bump and Tip luminosities are
presented in \S 4. Finally, in \S 5 we briefly summarize our results. 
  
\begin{table}
\begin{center}
\label{par}
\caption{Adopted parameters for the program clusters.}
\begin{tabular}{lcccc}
\hline\hline
\\
Cluster& [Fe/H]$_{CG97}$ &[M/H] & E(B\--V) & (m\--M)$_0$ \\
\\
\hline
\\
M 92 &-2.16  &-1.95 & 0.02 & 14.78\\
M 15& -2.12   & -1.91 & 0.09 &15.15\\
M 68 &-1.99  &  -1.81 & 0.04 & 15.14\\
M 30 &-1.91  &  -1.71 & 0.03 & 14.71 \\
M 55 &-1.61  &  -1.41 & 0.07 & 13.82 \\
${\omega}$~Cen & -1.60 & -1.39 & 0.11 & 13.70 \\
NGC 6752&-1.42  &  -1.21 & 0.04 &13.18 \\
M 10  &-1.41 &-1.25 & 0.28 & 13.38 \\
M 13 &-1.39  &-1.18 & 0.02 & 14.43 \\
M 3  &-1.34  &-1.16 & 0.01 & 15.03 \\
M 4  &-1.19  & -0.94  & 0.36 & 11.68 \\
NGC 362 &-1.15  &  -0.99 & 0.05 & 14.68 \\
M 5  &-1.11  &-0.90 & 0.03 & 14.37 \\
NGC 288 &-1.07  &  -0.85 & 0.03 & 14.73 \\
M 107 &-0.87 & -0.70 & 0.33 & 13.95 \\
NGC 6380 &-0.87 &  -0.68 & 1.29 & 14.81 \\
NGC 6342 &-0.71 &  -0.53 & 0.57 & 14.63 \\
47 Tuc&-0.70 & -0.59 & 0.04 & 13.32 \\
M 69 &-0.68  & -0.55 & 0.17 & 14.64 \\
NGC 6441 &-0.68 &  -0.52 & 0.52 & 15.65 \\
NGC 6624 &-0.63 &  -0.48 & 0.28 & 14.63 \\
NGC 6440&-0.49  &  -0.40 & 1.15 & 14.58 \\
NGC 6553&-0.44;-0.06$^*$ &  -0.36;-0.05$^*$ & 0.84 & 13.46 \\
NGC 6528 &-0.38;+0.07$^*$&  -0.31;+.05$^*$& 0.62 & 14.37\\
\\
\hline
\end{tabular}
\end{center}
(*) The most recent metallicity estimate by means of high resolution 
spectroscopy from \citet{eugi01}.
\end{table}
%
\section{The RGB Bump}
Theoretical models of stellar evolution predict that at some level along  
the RGB, the convective envelope penetrates deep enough into the star to reach 
the region of varying hydrogen abundance settled during the 
core hydrogen\--burning. When the convective envelope retreats from the advancing 
H\--burning shell, a discontinuity in the H abundance profile (X) is left. 
Thus, the advancing H\--burning shell {\it (i)} 
passes through the discontinuity and {\it (ii)} moves from a region of 
increasing X to a region of constant X. Event {\it (i)}  
generates a temporary drop in luminosity during the evolution of the star along
the RGB: from an observational point of view this evolutionary hesitation yields to a
peak in the differential luminosity function (LF). 
Event {\it (ii)} produces a change in the evolutionary rate of the star along the RGB, 
hence a change of the integrated LF slope. 
\citet{cr84} have made an extensive study of the best observable to locate the
RGB Bump, concluding that the integrated LF is a suitable tool to reliably
locate it.
\begin{figure}
\centering
\includegraphics[width=8.7cm]{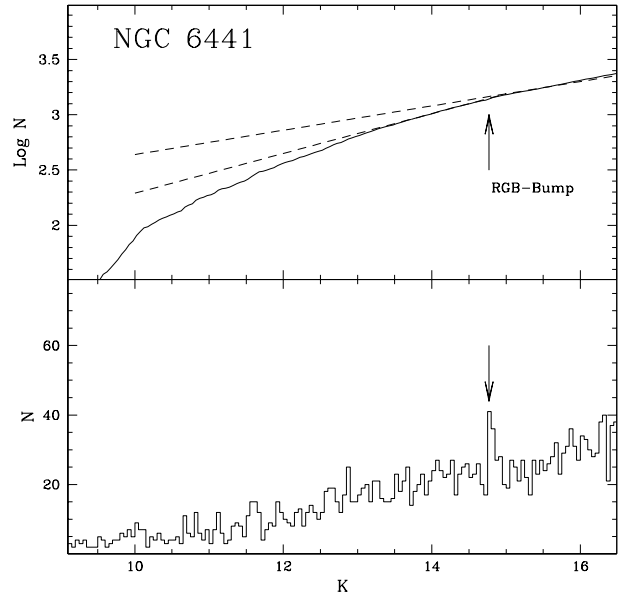}
\caption{Observed integrated (upper panel) and differential
(lower panel) luminosity function for RGB stars in NGC~6441. 
The dashed lines in the upper
panel are the linear fits to the region above and below the RGB Bump.}
\label{6441bump}
\end{figure}

\citet{fusi90} and more recently \citet{F99}, F00, 
\citet{zoc99,zoc00,coreani,riello,V04} and S04 
showed how this feature can be safely
identified in most of the current generation of optical and IR CMDs of GGCs,
and also in other stellar systems \citep[see e.g][]{bel01,bel02,lore02}. 
Basically, the main difficulty in detecting the RGB Bump 
is represented by the necessity of a large observed sample of RGB stars.
This becomes particularly difficult in the case of low\--metallicity GGCs,
where the Bump occurs in the brightest portion of the RGB, which is an 
intrinsically poorly populated sequence, due to the high\--evolutionary rate 
of stars at the very end of the RGB stage.
Following the same procedure as in F00 we identified the RGB Bump in
most of the clusters in our sample and in all the available photometric bands, 
by using both the differential and integrated LFs. As an example, 
Fig.~\ref{6441bump} shows both the LFs for the metal rich cluster NGC~6441.
In the case of M~30, our sample is not sufficiently large to reach a 
safely detection of the Bump. In order to increase the fraction of the 
sampled cluster population, complementary data from the 2MASS catalog have been 
also used. The RGB Bump identified from our data was re\--computed on the combined 
catalog. 
The inferred Bump K\--magnitude values can be transformed into V\--magnitudes 
by using the (V-K)$_0$ colours. The derived values are fully consistent 
with the direct determination of the Bump V\--magnitudes listed in Tab.~5 
of \citet{F99}.

\begin{table*}
\begin{minipage}{110mm}
\begin{center}
\caption{Near\--IR and Bolometric RGB Bump magnitudes for the observed clusters.}
\label{bump}
\begin{tabular}{l cc|c c c c|}
\hline\hline
\\
Name &[Fe/H]$_{CG97}$& [M/H] &J&H&K&M$_{Bol}$\\ 
\\
\hline
M 92 &-2.16  &-1.95 & 
12.85${\pm}0.05$&\---&12.35${\pm}0.05$ &-0.52${\pm}$0.21   \\
M 15& -2.12   & -1.91 & 
13.55${\pm}0.05$&13.05${\pm}0.05$&12.95${\pm}0.05$&-0.29${\pm}$0.21    \\
M 68 &-1.99  &  -1.81 & 
13.35${\pm}0.05$&\---&12.80${\pm}0.05$&-0.41${\pm}$0.21   \\
M 30 &-1.91  &  -1.71 & \---&\---&\---&\---\\
M 55 &-1.61  &  -1.41 & 
12.35${\pm}0.05$&\---&11.75${\pm}0.05$&-0.17${\pm}$0.21  \\
${\omega}$~Cen$^{(a)}$ &-1.60&-1.39&12.40${\pm}0.05$&\---&11.80${\pm}0.05$&
-0.02${\pm}$0.22  \\
NGC 6752&-1.42  &  -1.21 & 
11.90${\pm}0.05$&11.35${\pm}0.05$&11.25${\pm}0.05$&0.08${\pm}$0.21    \\
M 10  &-1.41 &-1.25 & \---&\---&\---&\---\\
M 13 &-1.39  &-1.18 & 
13.05${\pm}0.05$&\---&12.40${\pm}0.05$&0.02${\pm}$0.21  \\
M 3  &-1.34  &-1.16 & 
13.70${\pm}0.05$&\---&13.10${\pm}0.05$&0.13${\pm}$0.21  \\
M 4  &-1.19  & -0.94  &\---&\---&\---&\---\\
NGC 362 &-1.15  &  -0.99 & 
13.80${\pm}0.05$&13.25${\pm}0.05$&13.15${\pm}0.05$&0.53${\pm}$0.21  \\
M 5  &-1.11  &-0.90 & 
13.25${\pm}0.05$&\---&12.65${\pm}0.05$&0.33${\pm}$0.21  \\
NGC 288 &-1.07  &  -0.85 & 
13.75${\pm}0.05$&13.25${\pm}0.05$&13.20${\pm}0.05$&0.51${\pm}$0.21  \\
M 107 &-0.87 & -0.70 & 
13.25${\pm}0.05$&\---&12.50${\pm}0.05$&0.45${\pm}$0.21  \\
NGC 6380 &-0.87 &  -0.68 & 
15.15${\pm}0.05$&14.25${\pm}0.05$&13.95${\pm}0.05$&0.64${\pm}$0.22   \\
NGC 6342 &-0.71 &  -0.53 & 
14.65${\pm}0.10$&13.85${\pm}0.10$&13.75${\pm}0.10$&0.97${\pm}$0.22   \\
47 Tuc&-0.70 & -0.59 & 
12.65${\pm}0.05$&12.15${\pm}0.05$&12.05${\pm}0.05$&0.79${\pm}$0.21   \\
M 69 &-0.68  & -0.55 & 
14.25${\pm}0.05$&\---&13.53${\pm}0.05$&0.88${\pm}$0.21  \\
NGC 6441 &-0.68 &  -0.52 & 
15.70${\pm}0.05$&14.85${\pm}0.05$&14.77${\pm}0.05$&0.95${\pm}$0.22  \\
NGC 6624 &-0.63 &  -0.48 & 
14.45${\pm}0.05$&13.60${\pm}0.05$&13.65${\pm}0.05$&1.02${\pm}$0.22   \\
NGC 6440&-0.49  &  -0.40 & 
15.30${\pm}0.05$&14.35${\pm}0.05$&14.13${\pm}0.05$&1.21${\pm}$0.22  \\
NGC 6553&-0.44  &  -0.36 & 
14.05${\pm}0.1$&\---&13.05${\pm}0.10$&1.32${\pm}$0.22  \\
NGC 6528 &-0.38 &  -0.31 & 
15.10${\pm}0.1$&\---&14.05${\pm}0.10$&1.54${\pm}$0.22  \\
\hline
\multicolumn{7}{l}{(a) The observed RGB Bump magnitudes are from S04.}
\end{tabular}
\end{center}
\end{minipage}
\end{table*}
\begin{figure}
\centering
\includegraphics[width=8.7cm]{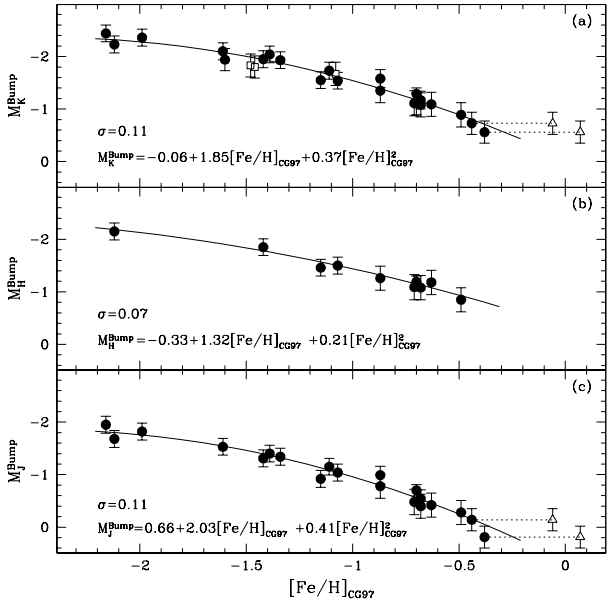}
\caption{RGB Bump absolute K (a), H (b) and J (c) 
magnitudes as a function of the cluster metallicity in the CG97 
scale. 
Filled circles: program clusters; empty squares: clusters in the 
\citet{coreani} dataset; empty triangles: NGC~6553 and NGC~6528 
with the most recent metallicity estimates from \citet{eugi01}. 
Solid lines are our best-fitting relations to the data.} 
\label{fe.bump}
\end{figure}
\begin{figure}
\centering
\includegraphics[width=8.7cm]{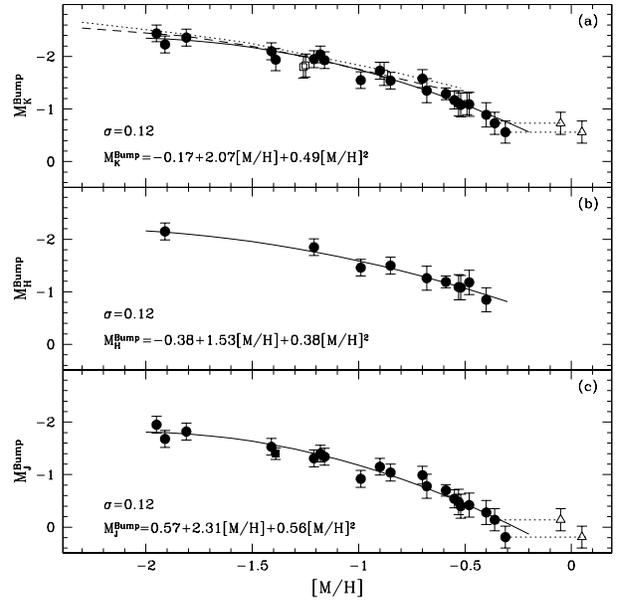}
\caption{
RGB Bump absolute K (a), H (b) and J (c)
magnitudes as a function of the cluster metallicity in the 
global metallicity scale ([M/H]).
Notation as in Fig.~\ref{fe.bump}.
Solid lines are our best-fitting relations to the data. 
The dotted and dashed lines in panel (a) are the theoretical 
predictions by SCL97 models at t=12 and 14 Gyr.}
\label{met.bump}
\end{figure}

The observed values of the RGB Bump for the global cluster sample 
are listed in Table~\ref{bump}. By adopting the reddening and distance moduli 
listed in Table~\ref{par} these values were converted 
into absolute magnitudes and their behaviour with varying the metallicity 
in the \citet[][ hereafter GC97]{CG97} ([Fe/H]$_{CG97}$) 
and in the global ([M/H]) scale defined by \citet{F99} have been shown in 
Figs.~\ref{fe.bump} and \ref{met.bump}.
By using 2MASS data, \citet{coreani} determined the M$_K$ Bump for 11
GGCs, 7 in common with our sample, spanning a quite large metallicity range. 
Their estimates are in nice
agreement with those of this work, F00, \citet{V04} and S04. 
In order to derive a more robust relation between the absolute M$_K$ magnitude 
and the metallicity, the best\--fitting relation has been obtained by using also
the \citet{coreani} data.
The comparison between the observational data and the theoretical predictions
based on the \citet[][ hereafter SCL97]{SCL97} models, 
shows an excellent agreement.
The inferred best\--fitting relations to the observed points 
(also reported in each panel of 
Figs.~\ref{fe.bump} and \ref{met.bump} with the corresponding standard deviation) 
are as follows: 

\begin{equation}
M^{Bump}_J=0.66+2.03[Fe/H]_{CG97}+0.41[Fe/H]^2_{CG97}
\end{equation}

\begin{equation}
M^{Bump}_H=-0.33+1.32[Fe/H]_{CG97}+0.21[Fe/H]^2_{CG97}
\end{equation}

\begin{equation}
M^{Bump}_K=-0.08+1.82[Fe/H]_{CG97}+0.36[Fe/H]^2_{CG97}
\end{equation}

\begin{equation}
M^{Bump}_J=0.57+2.31[M/H]+0.56[M/H]^2 
\end{equation}

\begin{equation}
M^{Bump}_H=-0.38+1.53[M/H]+0.38[M/H]^2 
\end{equation}

\begin{equation}
M^{Bump}_K=-0.17+2.07[M/H]+0.49[M/H]^2 
\end{equation}

\begin{figure}
\centering
\includegraphics[width=8.7cm]{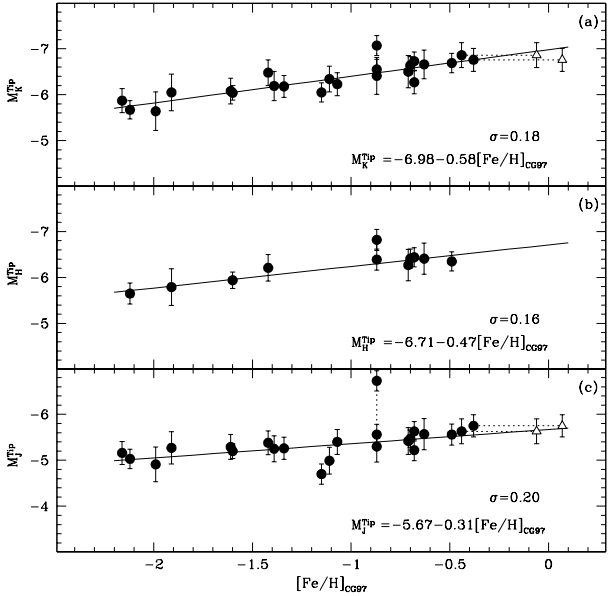}
\caption{J, H and K absolute magnitude of the RGB Tip as a function of the
metallicity in the CG97 scale for
the entire clusters sample.
The empty triangles refer to NGC~6553 and NGC~6528 with the most recent metallicity 
estimates by \citet{eugi01}.
Two points (filled circles) have been plotted for NGC~6380 
(see \S 4.6 for the discussion) and connected by a dotted line. 
The solid lines are our best\--fitting relations.} 
\label{fe.tip}
\end{figure}
\begin{figure}
\centering
\includegraphics[width=8.7cm]{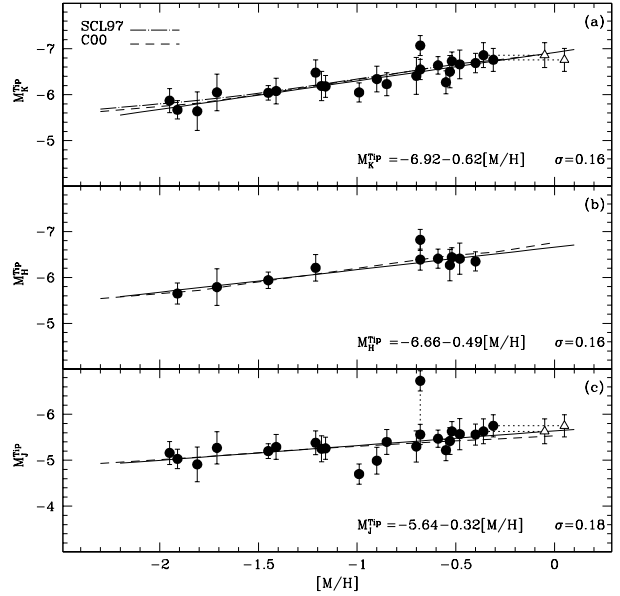}
\caption{J, H and K absolute magnitude of the RGB Tip as a function of the
global metallicity scale for the program clusters.
estimates by \citet{eugi01}.
Notation as in Fig.~\ref{fe.tip}.
The solid lines are our best\--fitting relations, the dashed lines 
are the theoretical predictions by C00 and the dot\--dashed line is the SCL97 
model.}
\label{met.tip}
\end{figure}

\section{The RGB Tip}
The evolution along the RGB ends at the so-called RGB Tip 
with the helium-ignition in the stellar core. 
In globular cluster stars this event is moderately violent (the so\--called
helium\--flash) because it takes place in a electron\--degenerate core. 
The luminosity of the RGB Tip is a quantity well
predicted by the theoretical models and, in recent years it has been widely
used as a standard candle to estimate distance to various stellar systems.
In fact, the RGB reaches its maximum extension in luminosity, 
for stellar populations older than ${\tau}{\approx}1-2$ Gyr 
(i.e. when stars less massive than $M{\approx}2.0 M_{\odot}$ are evolving) 
and it remains approximately constant increasing the age of the population. 
For this reason the method can be (in principle) successfully applied to
a variety of galaxies where a significant fraction of the population be
sufficiently old to have developed the full extension of the RGB. 
GGCs are the best template simple stellar populations where the RGB luminosity 
can be empirically calibrated as a function of metallicity. 
In this framework our group is currently performing
accurate empirical calibrations of the RGB Tip luminosity 
in various photometric passbands, in order to increase the
potentiality and the applicability of this method to the definition of the
distance scale.\\
Here we present a new calibration of the RGB Tip magnitudes with varying
metallicity in the near\--IR J, H and K bands. 
We adopt the method used in previous papers \citep[see F00 and][]{V04}, 
by assuming the brightest non\--variable star as representative of the RGB Tip 
level. Obviously, particular 
care was payed to decontaminate the sample from background field stars and,
more crucial, from the presence of Asymptotic Giant Branch (AGB) stars, 
which however, are significantly less numerous than RGB stars, 
given their much faster evolutionary rate. 
Moreover in low\--intermediate metallicity clusters
no long\--period AGB stars are expected. 

\begin{table*}
\begin{minipage}{110mm}
\caption{Observed and Bolometric RGB Tip magnitudes.}
\label{tip}
\begin{center}
\begin{tabular}{l cc|c c c c|}
\hline\hline
\\
Name &[Fe/H]$_{CG97}$& [M/H] &J&H&K&M$_{Bol}$\\ 
\\
\hline
M 92 &-2.16  &-1.95 & 
9.64${\pm}0.25$&\---&8.92${\pm}0.26$&-3.64${\pm}0.26$\\
M 15& -2.12   & -1.91 & 
10.20${\pm}0.21$&9.55${\pm}0.23$&9.42${\pm}0.20$&-3.55${\pm}0.20$\\
M 68 &-1.99  &  -1.81 & 
10.26${\pm}0.38$&\---&9.51${\pm}0.40$&-3.37${\pm}0.40$\\
M 30 &-1.91  &  -1.71 & 
9.45${\pm}0.29$&8.94${\pm}0.35$&8.67${\pm}0.35$&-3.70${\pm}0.35$\\
M 55 &-1.61  &  -1.41 & 
8.59${\pm}0.27$&\---&7.77${\pm}0.28$&-3.71${\pm}0.28$\\
${\omega}$~Cen$^{(a)}$ &-1.60&-1.39&8.59${\pm}$0.06&7.81${\pm}$0.08&
7.70${\pm}$0.06 &-3.59${\pm}0.16$  \\
NGC 6752$^{(b)}$&-1.42  &  -1.21 & 
7.84${\pm}0.25$&6.99${\pm}0.25$&6.72${\pm}0.28$&-3.65${\pm}0.28$\\
M 10  &-1.41 &-1.25 & \---&\---&\---&\--- \\
M 13 &-1.39  &-1.18 & 
9.20${\pm}0.28$&\---&8.25${\pm}0.32$&-3.59${\pm}0.32$\\
M 3  &-1.34  &-1.16 & 
9.78${\pm}0.24$&\---&8.85${\pm}0.24$&-3.61${\pm}0.24$\\
M 4  &-1.19  & -0.94  &\---&\---&\---&\---\\
NGC 362 &-1.15  &  -0.99 & 
10.02${\pm}0.22$&\---&8.65${\pm}0.21$&-2.90${\pm}0.21$\\
M 5$^{(c)}$&-1.11  &-0.90 & 
9.07${\pm}0.29$&\---&8.04${\pm}0.28$&-3.64${\pm}0.28$\\
NGC 288 &-1.07  &  -0.85 & 
9.36${\pm}0.25$&\---&8.51${\pm}0.25$&-3.80${\pm}0.25$\\
M 107 &-0.87 & -0.70 & 
8.94${\pm}0.34$&\---&7.67${\pm}0.40$&-3.57${\pm}0.40$\\
NGC 6380 &-0.87 &  -0.68 & 
10.37${\pm}0.22$&9.12${\pm}0.23$&8.75${\pm}0.22$&-3.88${\pm}0.22$\\
NGC 6342 &-0.71 &  -0.53 & 
9.71${\pm}0.29$&8.67${\pm}0.33$&8.35${\pm}0.32$&-3.70${\pm}0.32$\\
47 Tuc&-0.70 & -0.59 & 
7.88${\pm}0.19$&6.93${\pm}0.21$&6.69${\pm}0.19$&-3.71${\pm}0.19$\\
M 69 &-0.68  & -0.55 & 
9.57${\pm}0.23$&\---&8.43${\pm}0.25$&-3.51${\pm}0.25$ \\
NGC 6441 &-0.68 &  -0.52 & 
10.47${\pm}0.21$&9.49${\pm}0.21$&9.12${\pm}0.20$&-3.90${\pm}0.20$   \\
NGC 6624 &-0.63 &  -0.48 & 
9.30${\pm}0.32$&8.37${\pm}0.32$&8.08${\pm}0.31$&-3.85${\pm}0.31$  \\
NGC 6440&-0.49  &  -0.40 & 
10.02${\pm}0.23$&8.85${\pm}0.21$&8.33${\pm}0.21$&-3.82${\pm}0.21$    \\
NGC 6553&-0.44  &  -0.36 & 
8.56${\pm}0.27$&\---&6.92${\pm}0.27$&-3.86${\pm}0.27$    \\
NGC 6528 &-0.38 &  -0.31 & 
9.16${\pm}0.24$&\---&7.85${\pm}0.25$&-4.06${\pm}0.25$    \\
\hline
\multicolumn{7}{l}{(a) RGB Tip magnitudes from \citet{mic04}.}\\
\multicolumn{7}{l}{(b) RGB Tip magnitudes refer to 
star $19110813-6001517$ in the 2MASS catalog.}\\
\multicolumn{7}{l}{(c) RGB Tip magnitudes refer to star $15183604+0206373$ 
in the 2MASS catalog.} \\  
\end{tabular}
\end{center}
\end{minipage}
\end{table*}

The observed J, H and K magnitudes of the RGB Tip for the global sample are 
listed in Table~\ref{tip}. Fig.~\ref{fe.tip} and \ref{met.tip}
show the absolute RGB Tip magnitudes a function of the cluster metallicity 
in both the adopted scales. 
Two points have been plotted for NGC~6380. As discussed in Paper~I, its RGB
shape and its well\--defined HB clump suggest a metallicity higher than 47~Tuc,
thus it is expected to have several bright variable AGB
stars populating the upper part of the RGB, but none of them 
have been identified yet and a clear discrimination was not possible. 
Hence, since the brightest star could be a long\--period variable, we also 
consider as the possible {\it candidate} RGB Tip star 
the reddest among the brightest four stars in our photometry 
(the two filled circles connected by a dotted line in Fig.~\ref{fe.tip} and 
\ref{met.tip}).
Note that the region mapped by our observations covers the inner  
$4{\arcmin}\times4{\arcmin}$ in each cluster (see Paper~I), which 
allows us to sample a significant fraction of the cluster light (typically
${\approx}$30\%). However, in order to further check that we caught the brightest
RGB star, we also accurately inspect the CMD of the external
regions obtained from the 2MASS catalog. 
Due to its poor angular resolution (${\approx}2{\arcsec}$), the 2MASS survey is 
certainly not suitable to properly sample
the innermost regions of GGCs, but it can be successfully used to sample the
most external regions.
In two cases (namely M~5 and NGC~6752) a star in the 2MASS catalog brighter than
those sampled by our observations and lying along the cluster RGB ridge line 
has been found.
Thus, for these two clusters we assumed the magnitudes of the 2MASS stars
as best estimate of the RGB Tip.
Note that in the case of M~5, the new estimate found here replaces the previous
estimate by \citet{V04}.\\
Figs.~\ref{fe.tip} and \ref{met.tip} also report the RGB Tip
determination for the dominant population of ${\omega}$~Cen
recently obtained by \citet{mic04}. 
${\omega}$~Cen is the most massive stellar system 
in the Galactic halo (its mass is ${\approx}1$ order of magnitude larger
than the one of {\it normal clusters}), and its RGB Tip level
was measured from the sharp cut\--off of the RGB LF detected by applying the
edge\--detector filter (the so\--called Sobel filter). In order to
include this measure in our sample, we adopt the metallicity of the dominant
population ([Fe/H]$_{CG97}$=-1.60, [M/H]=-1.39), the distance modulus
((m-M)$_0$=13.70) and the reddening (E(B\--V)=0.11), as done 
by \citet{mic04}.\\ 
The inferred best\--fitting relations to the observed points 
(also reported in each panel of 
Figs.~\ref{fe.tip} and \ref{met.tip} with the corresponding standard deviation) 
are as follows: 
\begin{equation}
M^{Tip}_J=-5.67-0.31[Fe/H]_{CG97}
\end{equation}

\begin{equation}
M^{Tip}_H=-6.71-0.47[Fe/H]_{CG97}
\end{equation}

\begin{equation}
M^{Tip}_K=-6.98-0.58[Fe/H]_{CG97}
\end{equation}

\begin{equation}
M^{Tip}_J=-5.64-0.32[M/H]
\end{equation}

\begin{equation}
M^{Tip}_H=-6.66-0.49[M/H]
\end{equation}

\begin{equation}
M^{Tip}_K=-6.92-0.62[M/H]
\end{equation}

Statistical fluctuations are the main source of uncertainty, 
since the upper region of the RGB is intrinsically poorly populated. 
Following F00, we computed the expected error on the basis of the number
of stars in the brightest two magnitude bin along the RGB.
For the program clusters, the ${\sigma}_{stat}$ ranges from 0.03, 
for the most populated up to 0.32 for the least populated clusters. 
In order to minimize the
statistical fluctuations, complementary data from 2MASS were also used and the
${\sigma}_{stat}$ of our data was re\--computed on the combined sample. The
derived ${\sigma}_{stat}$ value have been considered as representative 
of the main uncertainty in the determination of the RGB Tip.
Of course, the accuracy of the RGB Tip estimate in ${\omega}$~Cen 
(0.16, 0.18 and 0.16 mag in J, H and K bands, respectively, see \citet{mic04})
is by far significantly higher than that obtained in the other clusters. 
As can be seen from Fig.~\ref{met.tip}, our derived relations and
the theoretical predictions by \citet[][ hereafter COO]{C00}
and by SCL97 are in excellent agreement in all the three bands.

\begin{figure}
\centering
\includegraphics[width=8.7cm]{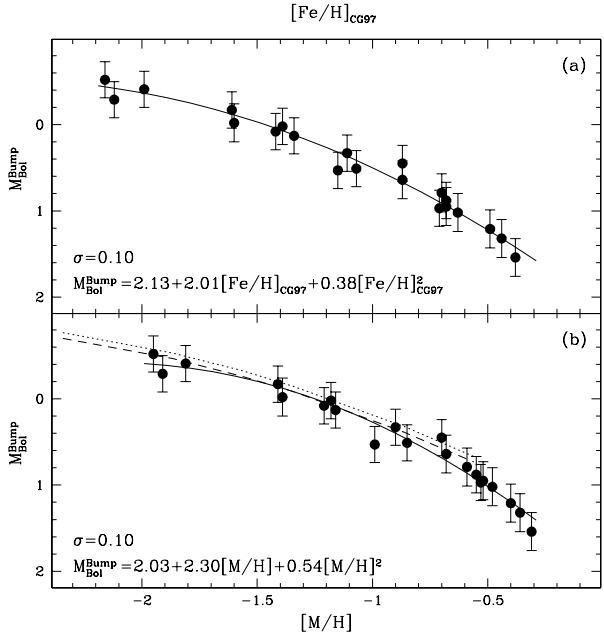}
\caption{Bolometric magnitudes of the RGB Bump as
a function of the cluster metallicity in both the adopted metallicity 
([Fe/H]$_{CG97}$\--panel (a) and [M/H]\--panel (b)) scales. 
The solid lines
are our best\--fitting relations, the dashed line in panel (b) is the
theoretical prediction by SCL97 models at t=12 and 14 Gyr.}
\label{bol_bump}
\end{figure}
\begin{figure}
\centering
\includegraphics[width=8.7cm]{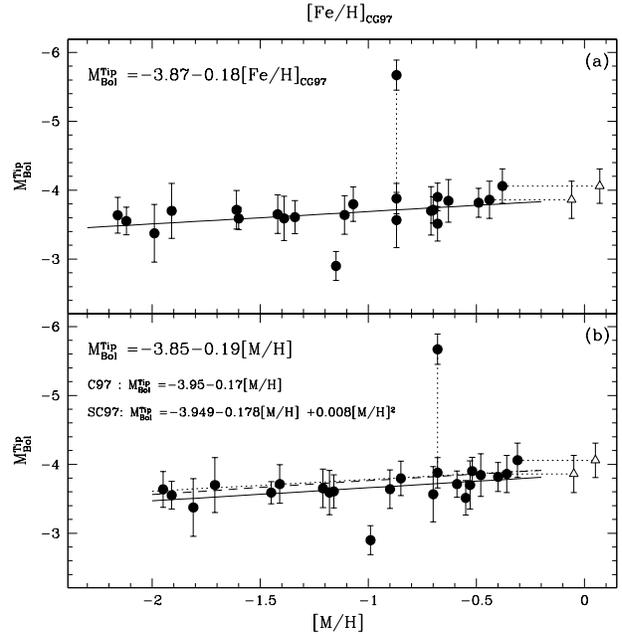}
\caption{Bolometric magnitudes of the RGB Tip as
a function of the cluster metallicity in both the adopted  
metallicity ([Fe/H]$_{CG97}$\--panel (a) and [M/H]\--panel (b)) scales. 
The solid lines are our best\--fitting relations. Two theoretical
predictions have been plotted in panel (b): 
\citet[][ C97]{C97} (dotted line) and
\citet[][ SC97]{SC97} (dashed line).}
\label{bol_tip}
\end{figure}

\section{The theoretical plane}
Accordingly to F00, the magnitudes of the RGB Bump and Tip were transformed 
in the theoretical plane by using the bolometric corrections for Population~II 
giants computed by \citet{paolo98}.
Fig.~\ref{bol_bump} shows the bolometric magnitudes of the RGB Bump for the 
program clusters. Two quadratic relations 
giving RGB Bump bolometric magnitude as a 
function of both the adopted metallicity scales have been derived:

\begin{equation}
M^{Bump}_{Bol}=2.13+2.01[Fe/H]_{CG97}+0.38[Fe/H]^2_{CG97}
\end{equation}

\begin{equation}
M^{Bump}_{Bol}=2.03+2.30[M/H]+0.54[M/H]^2
\end{equation}

The comparison  between the empirical estimates and the theoretical predictions
based on the SCL97 models (dashed line in panel [b] of Fig.~\ref{bol_bump}), 
show an excellent agreement.
Finally, Fig.~\ref{bol_tip} shows the bolometric magnitudes of the RGB Tip for 
the program clusters with varying metallicity in both the CG97 and global scales. 
The best\--fitting relations to our data (solid line in Fig.~\ref{bol_tip}) 
are:

\begin{equation}
M^{Tip}_{Bol}=-3.87-0.18[Fe/H]_{CG97}
\end{equation}

\begin{equation}
M^{Tip}_{Bol}=-3.85-0.19[M/H]
\end{equation}

Two theoretical relations have been overplotted in
panel (b) of Fig.~\ref{bol_tip}: \citet{C97} (dotted line) and \citet{SC97} 
(dashed line), respectively. Both models 
nicely agree with observations. As remarked by F00, 
the theoretical models have to be considered as an upper luminosity boundary of
the observed estimates because of the statistical fluctuations affecting 
the observed RGB samples, intrinsically poorly populated in globular clusters 
\citep{cast}.

\section{Conclusions}
New calibrations of the RGB Bump and Tip in the J, H and K bands 
as well as in bolometric , 
based on a global sample of 24 GGCs, have been presented. 
The behaviour of these evolutionary features have been 
investigated with varying the cluster metallicity in the CG97 and 
global scale, thus taking into account the effect of the 
${\alpha}$\--enhancement.\\
Quadratic and linear best-fitting relations linking the RGB Bump and Tip 
magnitudes and metallicity, respectively, have been obtained 
(see eqs. 1-16).  
Comparisons between observations and theoretical models show 
a good agreement.\\
The RGB Bump and Tip represent powerful tools to obtain 
independent estimates of metallicity and distance, respectively,  
in old stellar systems within the Local Group.
The new IR adaptive optics facilities available at ground-based 8m-class 
telescopes as well as the future imaging capabilities of the 
James Webb Space Telescope are allowing to use the RGB Tip distance indicator  
in galaxies well beyond the Local Group, up to a few Mpc away.

%

\section*{Acknowledgments}
The financial support by the Agenzia Spaziale Italiana (ASI) and the
Ministero dell'Istruzione, Universit\'a e Ricerca (MIUR) is kindly
acknowledged.
Part of the data analysis has been performed with the software developed by P.
Montegriffo at the INAF\--Osservatorio Astronomico di Bologna.
This publication makes use of data products from the Two Micron All Sky Survey,
which is a joint project of the University of Massachusetts and Infrared
Processing and Analysis Center/California Institute of Technology, founded by
the National Aeronautics and Space Administration and the National Science
Foundation.

\label{lastpage}


\begin{thebibliography}{199}
\bibitem[\protect\citeauthoryear{Bellazzini et al.}{2001b}]{bel01}
Bellazzini, M., Ferraro, F.~R. \& Pancino, E. 2001, MNRAS, 327,15

\bibitem[\protect\citeauthoryear{Bellazzini et al.}{2002}]{bel02}
Bellazzini, M., Ferraro, F.~R., Origlia, L., Pancino, E., Monaco, L. \& Oliva,
E. 2002, AJ, 124,3222

\bibitem[\protect\citeauthoryear{Bellazzini et al.}{2004}]{mic04}
Bellazzini, M., Ferraro, F.~R., Sollima, A., Pancino, E. \& Origlia, L. 2004,
A\&A, submitted

\bibitem[\protect\citeauthoryear{Caloi et al.}{1997}]{C97}
Caloi, V., D'Antona, F. \& Mazzitelli, I. 1997, A\&A, 320, 823 (C97)

\bibitem[\protect\citeauthoryear{Carretta \& Gratton}{1997}]
{CG97}
Carretta, E. \& Gratton, R.~G. 1997, A\&AS, 12, 95 (CG97)

\bibitem[\protect\citeauthoryear{Carretta et al.}{2001}]{eugi01}
Carretta, E., Cohen, J.~G. \& Gratton, R.~G. 2001, AJ, 122,1469

\bibitem[\protect\citeauthoryear{Cassisi et al.}{2000}]{C00}
Cassisi, S., Castellani, V., Ciarcelluti, P., Piotto, G. \& Zoccali, M. 2000,
MNRAS, 315,679 (C00)

\bibitem[\protect\citeauthoryear{Castellani, Degl'Innocenti \& Luridiana}{1993}]{cast}
Castellani, V, Degl'Innocenti, S. \& Luridiana, V. 1993, A\&A, 272,558

\bibitem[\protect\citeauthoryear{Cho \& Lee}{2002}]{coreani}
Cho, D.~H. \& Lee, S.~G. 2002, AJ, 124,977

\bibitem[\protect\citeauthoryear{Crocker \& Rood}{1990}]{cr84}
Crocker, D.~A. \& Rood, R.~T. 1984, in The Observational Tests of the Stellar
Evolution Theory, ed. A. Maeder \& A. Renzini (Dordrecht: Reidel), 159

\bibitem[\protect\citeauthoryear{Ferraro et al.}{1999}]{F99}
Ferraro, F.~R., Messineo, Fusi Pecci, F., De Palo, M.~A., Straniero, O., 
Chieffi, A. \& Limongi, M.
1999, AJ, 118, 1738 (F99)

\bibitem[\protect\citeauthoryear{Ferraro et al.}{2000}]{F00}
Ferraro, F.~R., Montegriffo, P., Origlia, L., \& 
Fusi Pecci, F. 2000, AJ, 119, 1282, (F00)

\bibitem[\protect\citeauthoryear{Fusi Pecci et al.}{1990}]{fusi90}
Fusi Pecci, F., Ferraro, F.~R., Crocker, D.~A., Rodd, T.~R. \& Buonanno, R.
1990, A\&A, 238,95

\bibitem[\protect\citeauthoryear{Iben}{1968}]{iben68}
Iben, I.~Jr. 1968, Nature, 220,143

\bibitem[\protect\citeauthoryear{Monaco et al.}{2002}]{lore02}
Monaco, L., Ferraro, F.~R., Bellazzini, M. \& Pancino, E. 2002, AJ, 578,50

\bibitem[\protect\citeauthoryear{Montegriffo et al.}{1998}]{paolo98}
Montegriffo, P., Ferraro, F.~R., Origlia, L. \& Fusi Pecci, F. 1998, MNRAS,
297, 872

\bibitem[\protect\citeauthoryear{Moorwood et al.}{1992}]{irac2}
Moorwood et al. 1992, The Messenger, 69,61

\bibitem[\protect\citeauthoryear{Riello et al.}{2003}]{riello}
Riello, M., Cassisi, S., Piotto, G., De~Angeli, F., Salaris, M., 
Pietrinferni, A., Bono, G. \& Zoccali, M. 2003, A\&A, 410, 553

\bibitem[\protect\citeauthoryear{Salaris \& Cassisi}{1997}]{SC97}
Salaris, M. \& Cassisi, S. 1997, MNRAS, 289, 406 (SC97)

\bibitem[\protect\citeauthoryear{Savage \& Mathis}{1979}]{SavMat}
Savage, B.~D. \& Mathis, J.~S. 1979, ARA\&A, 17, 73

\bibitem[\protect\citeauthoryear{Sollima et al.}{2004}]{BB}
Sollima, A., Ferraro, F.~R., Origlia, L., Pancino, E. \& Bellazzini, M. 2004,
A\&A astrp-ph/0402100, in press (S04) 

\bibitem[\protect\citeauthoryear{Straniero et al.}{1997}]{SCL97}
Straniero, O., Chieffi, A. \& Limongi, M. 1997, ApJ, 490,425 (SCL97)
 
\bibitem[\protect\citeauthoryear{Valenti et al.}{2004a}]{V04}
Valenti, E., Ferraro, F.~R., Perina, S. \& Origlia, L. 2004, A\&A, 
astro-ph/0401153, in press 

\bibitem[\protect\citeauthoryear{Valenti et al.}{2004b}]{pap1}
Valenti, E., Ferraro, F.~R. \& Origlia, L. 2004, MNRAS, 
astro-ph/0403536, in press (Paper I)

\bibitem[\protect\citeauthoryear{Zoccali et al.}{1999}]{zoc99}
Zoccali, M., Cassisi, S., Piotto, G., Bono, G. \& Salaris, M. 1999, AJ, 518, 49

\bibitem[\protect\citeauthoryear{Zoccali et al.}{2000}]{zoc00}
Zoccali, M. \& Piotto, G. 2000, A\&A, 358,943

\end{thebibliography}
\end{document}